\documentclass[%
 aip,
 amsmath,amssymb,
 reprint,%
]{revtex4-1}

\usepackage{graphicx}% Include figure files
\usepackage{dcolumn}% Align table columns on decimal point
\usepackage{bm}% bold math
\usepackage[utf8]{inputenc}
\usepackage[T1]{fontenc}
\usepackage{mathptmx}
\usepackage{etoolbox}
\usepackage{hyperref} 
\hypersetup{colorlinks,citecolor=blue, filecolor=blue ,linkcolor=blue , urlcolor=blue, pdftex}
\usepackage[dvipsnames]{xcolor}

\def \FUW{Institute of Experimental Physics, Faculty of Physics, University of Warsaw, ul. Pasteura 5, 02-093 Warsaw, Poland}

\def \Watanabe{Research Center for Functional Materials, National Institute for Materials Science, 1-1 Namiki, Tsukuba 305-0044, Japan}
\def \Taniguchi{International Center for Materials Nanoarchitectonics, National Institute for Materials Science, 1-1 Namiki, Tsukuba 305-0044, Japan}

\makeatletter
\def\@email#1#2{%
 \endgroup
 \patchcmd{\titleblock@produce}
  {\frontmatter@RRAPformat}
  {\frontmatter@RRAPformat{\produce@RRAP{*#1\href{mailto:#2}{#2}}}\frontmatter@RRAPformat}
  {}{}
}%
\makeatother
\begin{document}

\preprint{AIP/123-QED}

\title[]{The effect of dielectric environment on the brightening of neutral and charged dark excitons in WSe$_2$ monolayer}
% Force line breaks with \\
\author{Małgorzata Zinkiewicz}
\affiliation{\FUW}
\author{Magdalena Grzeszczyk}
\affiliation{\FUW}
\author{Łucja Kipczak}
\affiliation{\FUW}
\author{Tomasz~Kazimierczuk}
\affiliation{\FUW}
\author{Kenji~Watanabe}
\affiliation{\Watanabe}
\author{Takashi Taniguchi}
\affiliation{\Taniguchi}
\author{Piotr Kossacki}
\affiliation{\FUW}
\author{Adam Babi\'nski}
\affiliation{\FUW}
\author{Maciej R. Molas}
\email{maciej.molas@fuw.edu.pl}
\affiliation{\FUW}

\begin{abstract}
The dielectric environment of atomically-thin monolayer (ML) of semiconducting transition metal dichalcogenides affects both the electronic band gap and the excitonic binding energy in the ML. We investigate the effect of the environment on the in-plane magnetic field brightening of neutral and charged dark exciton emissions in the WSe$_2$ ML. The monolayers placed in three dielectric environments are studied, in particular, the ML encapsulated in hexagonal BN (hBN) flakes, the ML deposited on a hBN layer, and the ML embedded between the hBN flake and SiO$_2$/Si substrate. We observe that the brightening rates of the neutral and charged dark excitons depend on the dielectric environment, which may be related to the variation of the level of carrier concentration in the ML. Moreover, the surrounding media, characterized by different dielectric constants, influences weakly the relative energies of the neutral and charged dark excitons in reference to the bright ones.
\end{abstract}

\maketitle

%XXXXXXXXXXXXXXXXXXXXXXXX        INTRO
%\section{Introduction \label{sec:Intro}}
Two-dimensional (2D) semiconducting transition metal dichalcogenides (S-TMDs) have attracted significant attention in the last decade due to their thickness-dependent electronic band structure as well as intriguing physics occurring in the monolayer (ML) limit~\cite{Koperski2017, Wang2018}. 
Particularly, the binding energy of neutral excitons in S-TMD MLs can be modified significantly from about 290~meV for a ML placed on SiO$_2$/Si substrate~\cite{Raja2017} to about 170~meV for the ML encapsulated in hexagonal BN (hBN)~\cite{Molas2019,Goryca2019}, which results from the nonuniform dielectric screening of excitonic states. 
In contrast, the separation energy between the neutral and charged excitons, known as the trion binding energy, is only weakly affected by the surrounding dielectric media~\cite{Courtade2017,Vaclavkova2018}. 
To complete this picture, we investigate the effect of the environment on optical properties of dark (optically inactive) excitons. 
This is important especially in tungsten-based MLs, $i.e.$ WSe$_2$ and WS$_2$, as their low-temperature photoluminescence (PL) spectra are dominated by the emission lines associated with recombination processes involving dark excitons. 

We study the brightening of neutral and charged dark excitons in the WSe$_2$ ML by means of the in-plane magnetic field.  
The MLs in three dielectric environments are studied: (i) the ML encapsulated in hBN flakes, (ii) the ML deposited on hBN layer, (iii) the ML embedded between the hBN flake and the SiO$_2$/Si substrate. 
The brightening of dark excitons is found to depend on the dielectric environment. This effect is ascribed to the variation of the carrier concentration in the ML embedded between different surrounding media. 
Our results devoted to the relative energies of the neutral and charged dark excitons in reference to the bright ones were compared with those reported in the literature. 
A weak influence of the different dielectric environment on those energies was observed.

%\section{Results and Discussion \label{results}}
\begin{figure}[h!b]
%		\subfloat{}%
		\centering
		\includegraphics[width=1\linewidth]{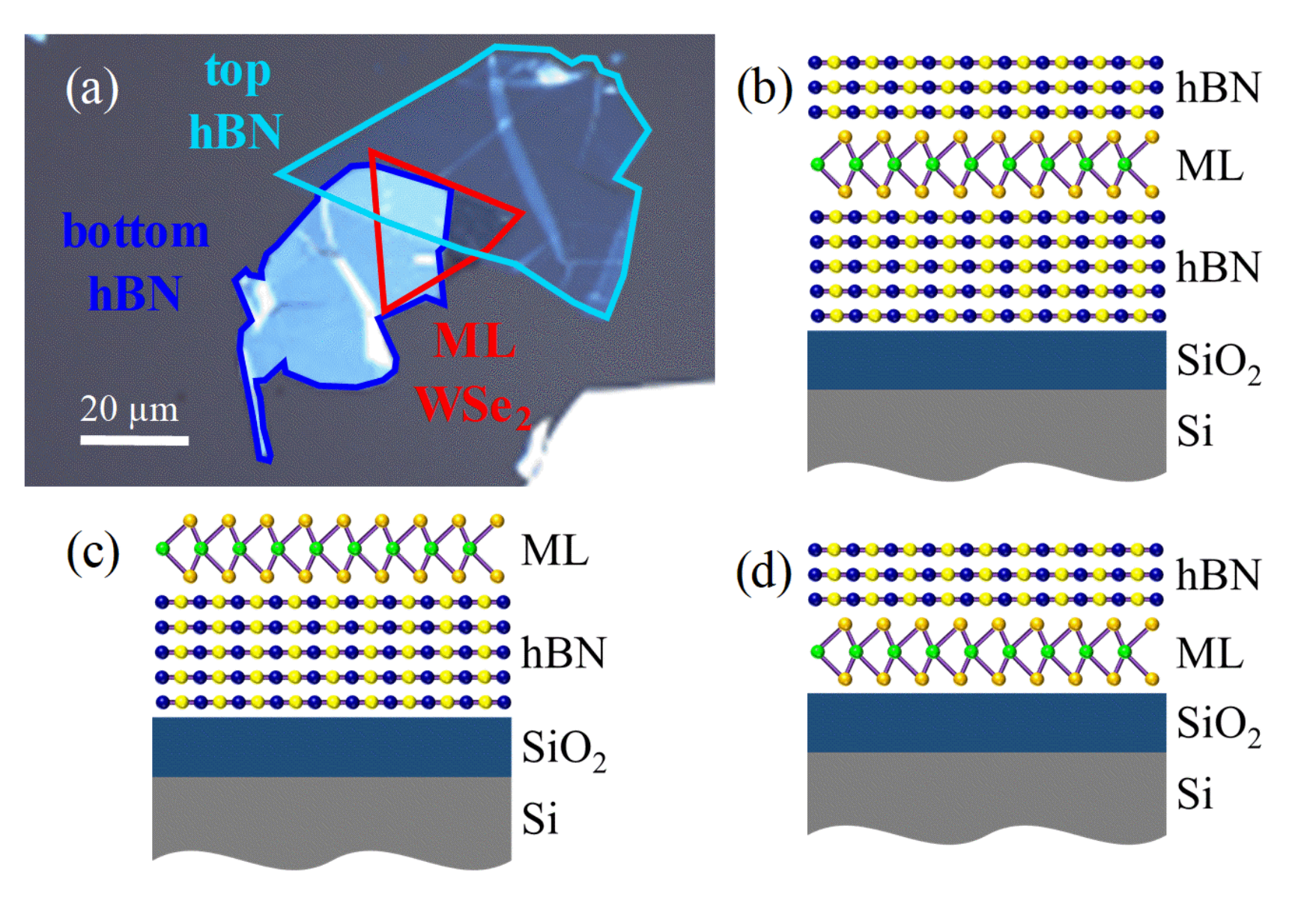}%
    	\caption{(a) Optical image of the sample under study. The schematic illustration of the WSe$_2$ monolayers (b) encapsulated in hBN flakes, (c) deposited on bottom hBN flake, and (d) deposited on SiO$_2$/Si substrate and covered with top hBN flake.}
		\label{fig:fig0}
\end{figure}

The investigated sample is composed of  a WSe$_2$ ML and hBN layers that were fabricated by two-stage PDMS (polydimethylsiloxane)-based mechanical exfoliation~\cite{Gomez}. 
The WSe$_2$ ML flake was placed on the SiO$_2$(90~nm)/Si substrate, which was partially covered with a thick bottom hBN layer. 
The heterostructure was partially capped with a thin top hBN flake. 
As a result, three regions of the samples could be identified: hBN/WSe$_2$/hBN/SiO$_2$/Si, WSe$_2$/hBN/SiO$_2$/Si, and hBN/WSe$_2$/SiO$_2$/Si. 
The optical image of the sample and the schematic representations of different sample areas are shown in Fig.~\ref{fig:fig0}.

\begin{figure*}[h!t]
%		\subfloat{}%
		\centering
		\includegraphics[width=0.9\linewidth]{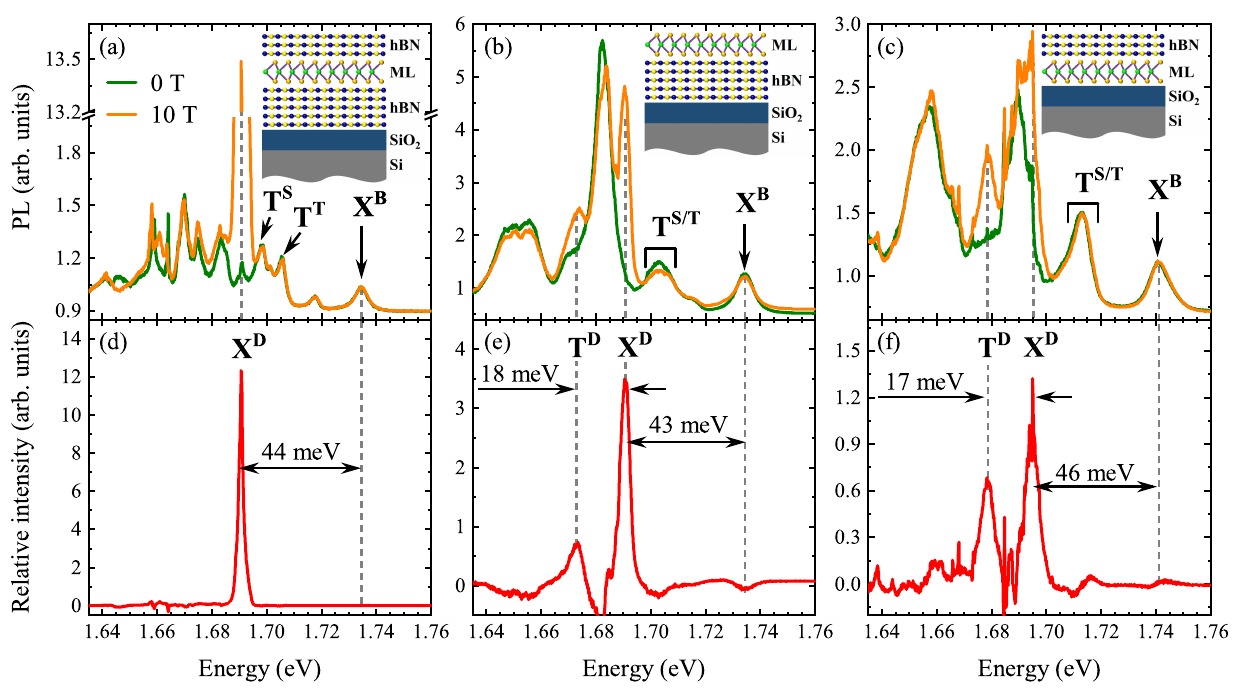}%
    	\caption{(Upper panels) PL spectra of WSe$_2$ monolayer: (a) encapsulated in hBN flakes, (b) deposited on hBN flake, and (c) deposited on SiO$_2$/Si substrate and covered with hBN flake at $T$=10~K measured at zero field (green curves) and at $B$=10~T (orange curves) applied in the plane of the crystal. The PL spectra were normalized to the intensity of the bright X$^\textrm{D}$ line. (Lower panels) Corresponding relative intensities of MLs defined as $\textrm{PL}_{B=10~\textrm{T}}-\textrm{PL}_{B=0~\textrm{T}}$ are represented by red curves.}
		\label{fig:fig1}
\end{figure*}

Micro-magneto-PL experiments were performed in the Voigt configuration using a free-beam-optics arrangement in a superconducting coil producing magnetic fields up to 10~T. The samples were placed in a helium gas atmosphere at temperature of 10~K. 
The excitation light was focused by a single aspheric lens to a spot smaller than 2 $\mu$m. 
The lens was mounted on the x-y-z piezoelectric stage inside the cryostat. 
The excitation light was provided by a $\lambda$=532 nm (2.33 eV) continuous wave (CW) laser diode. 
The emitted light was dispersed with a 0.5 m long monochromator and detected with a charge-coupled device (CCD) camera. 

It is well established in the literature~\cite{Molas2017, Zhang2017, Molas2019, Lu2019, Robert2020, Feierabend2020} that the application of the in-plane magnetic field leads to the brightening of both the charged and neutral dark excitons in MLs of S-TMDs. 
To investigate the effect of an in-plane magnetic field on the low temperature ($T$=10~K) PL of the WSe$_2$ ML placed in the different dielectric environments, we measured the evolution of the corresponding PL spectra in the Voigt configuration as a function of an external magnetic field up to $B$=10~T. 

The PL spectra at $B$=0~T and at $B$=10~T are presented in the upper panels of Fig.~\ref{fig:fig1}. 
The zero-field PL spectra of all measured areas of the ML display a characteristic emission line, labeled X$^\textrm{B}$, associated with recombination of the neutral bright excitons formed at the K$^\pm$ points of the Brillouin zone~\cite{Koperski2017}. 
As can be seen in the Figure, the zero-field PL spectra, apart from the X$^\textrm{B}$ peaks, consist of several emission lines on the lower energy side of the spectrum, which energies and linewidths are significantly modified on different areas of the sample. 
These additional lines have been attributed in the literature to the charged excitons (trions), neutral and charged biexcitons, phonon replicas, localized excitons, and so on~\cite{Arora2015, Smolenski2016, Molas2017, Zhang2017, Robert2017, Wang2017, Barbone2018, Chen2018, Li2018, Koperski2019, Paur2019, Molas2019, LiuGate, Li2019, arora2019dark, LiuValley, Liu2020, He2020, Sven2020,jadczak2021, yang2022}, described in detail in the Supplementary Material (SM). 
The application of magnetic field in the plane of the ML strongly affects the measured PL spectra at energies 40-60~meV below the X$^\textrm{B}$ line. 
The corresponding PL spectra of the three investigated areas are shown in the SM.
To better visualize the effect and compare the results obtained for different sample areas, we define a relative spectrum in the magnetic field as $\textrm{PL}_{B\neq 0}-\textrm{PL}_{B=0}$. 
The relative spectra for $B$=10~T are shown in the lower panels of Fig.~\ref{fig:fig1}. 
In the case of WSe$_2$ ML encapsulated in hBN, the spectrum comprises a single narrow line, labeled X$^\textrm{D}$, which we ascribed to the neutral dark exciton~\cite{Robert2017, Wang2017, Molas2019}. 
This complex in the K$^\pm$ valley of the Brillouin zone is composed of an electron from the lowest-lying spin-polarized level of the conduction band and a hole from the highest-lying spin-polarized valence band, see Ref.~\cite{Molas2017,Molas2019} for details. 
%Note that the observed asymmetric shape of this peak is related to the appearance of the fine structure of the neutral dark exciton~\cite{Robert2017, Molas2019}. 
For WSe$_2$ ML deposited on the hBN flake or deposited on the SiO$_2$/Si substrate and covered with the hBN flake, the corresponding spectra are composed of two peaks, labeled X$^\textrm{D}$ and T$^\textrm{D}$. 
The latter peak is associated with the recombination of the negative dark trion\cite{arora2019dark,LiuGate}. 
The negative dark trion at the K$^\pm$ point is formed by the energetically lowest electron-hole pair at K$^\pm$ point and an extra electron located in the lowest-lying level in the opposite K$^\mp$.
Note that the negative sign of the free carriers in the investigated monolayers is determined
by the observation of fine structure of the trion line for the WSe$_2$ ML encapsulated in hBN,
$i.e.$ clearly resolved lines related to the intervalley spin-singlet (T$^\textrm{S}$) and intervalley spin-triplet (T$^\textrm{T}$) negative trions~\cite{Courtade2017, He2020}.
Due to increased linewidths, this splitting cannot be readily resolved for the two other MLs, the corresponding emission line is denoted as T$^\textrm{S/T}$,
but the sole modification of the dielectric environment cannot change the sign of majority carriers (see Ref.~\citenum{Grzeszczyk2021} for details).

\begin{figure}[t]
	%	\subfloat{}%
		\centering
		\includegraphics[width=1.0\linewidth]{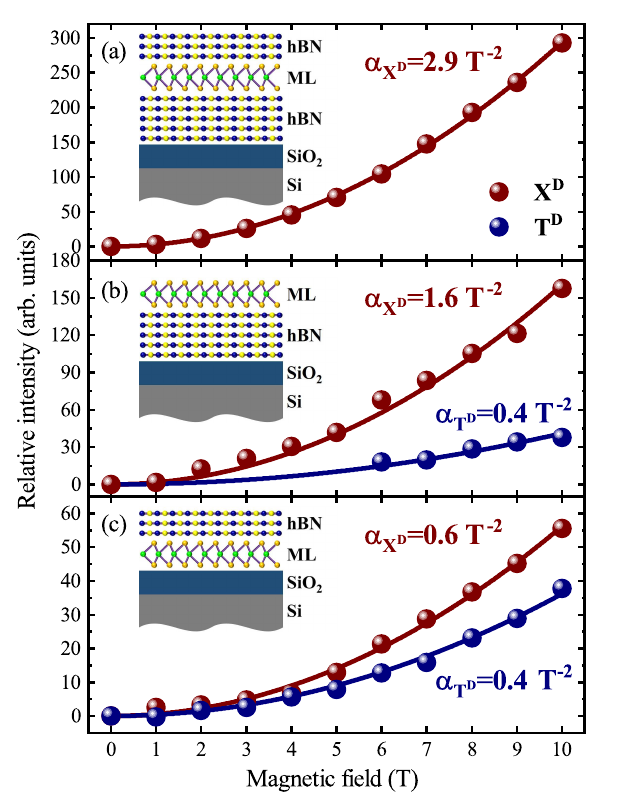}%
    	\caption{Magnetic-field dependence of the intensities of dark exciton and trion lines obtained on the WSe$_2$ monolayers: (a) encapsulated in hBN flakes, (b) deposited on hBN flake, and (c) deposited on SiO$_2$/Si substrate and covered with hBN flake. The solid red and blue curves represent quadratic fits.}
		\label{fig:fig2}
\end{figure}

To further verify the assignment of the X$^\textrm{D}$ and T$^\textrm{D}$ lines, we investigate their integrated intensities ($I$) as a function of in-plane magnetic field.
The dependence is expected to be quadratic $I=\alpha B^2$ (see Ref.~\citenum{Molas2017} for details).
The fitting parameter $\alpha$ will be referred to as the brightening coefficient. 
The experimental data accompanied with the fitted curves are presented in Fig.~\ref{fig:fig2}. 
The obtained $\alpha_{\textrm{X}^\textrm{D}}$ coefficients decrease significantly from the value of $2.9~\textrm{T}^{-2}$ for the encapsulated ML, through $1.6~\textrm{T}^{-2}$ for the ML deposited on hBN, and ending with $0.6~\textrm{T}^{-2}$ for the ML embedded between the hBN flake and SiO$_2$/Si substrate. 
At the same time, there is no signature of the dark trion emission in the PL spectra measured on the area encapsulated in hBN. 
The corresponding $\alpha_{\textrm{T}^\textrm{D}}$ coefficients attain comparable values ($0.4~\textrm{T}^{-2}$) for two other investigated ML areas. 
The observed effects of the dielectric environment on the brightening of both the neutral and charged dark excitons may be explained in terms of the doping level of the corresponding area. 
The doping level for the WSe$_2$ ML encapsulated in hBN has been found to be close to the neutrality point, which favors the creation of the neutral dark excitons~\cite{Molas2019}. 
Oppositely, the brightening of the negative dark trions has been facilitated in the $n$-doped WS$_2$ ML encapsulated in hBN~\cite{Zinkiewicz2020, Zinkiewicz2021}. 
It is also known that hBN layers serve as barriers preventing charge transfer from impurities $e.g.$ those present in SiO$_2$/Si substrates~\cite{Illarionov2016}, which may lead to the modification of carrier concentration~\cite{Grzeszczyk2021}. 
We can conclude that almost the same brightening coefficients for both the neutral and charged dark excitons for the WSe$_2$ ML deposited on SiO$_2$/Si substrate and covered with hBN are associated with the highest doping level. 
The doping decrease for the ML deposited on bottom hBN, leaving the ML encapsulated in hBN close to the neutrality point. 
Note that the additional analysis of the doping level is presented in the SM.

\begin{figure}[b!]
	%	\subfloat{}%
		\centering
		\includegraphics[width=1.0\linewidth]{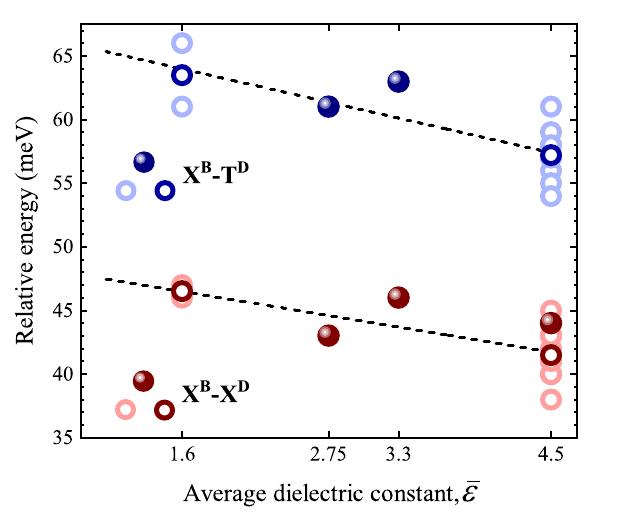}%
    	\caption{The dependence of the relative energies, defined as the energy difference of the X$^\textrm{B}$ and X$^\textrm{D}$/T$^\textrm{D}$ lines. 
    	The full points correspond to the data presented in Fig.~\ref{fig:fig1}, while the open bright points represent the data from Refs.~\citenum{Molas2017, Zhang2017, Robert2017, Wang2017, Barbone2018, Chen2018, Li2018, Paur2019, Molas2019, LiuGate, Li2019, arora2019dark, LiuValley, Liu2020, He2020, Sven2020, jadczak2021, yang2022}. The open dark points denote the calculated mean values based on the presented data. The dashed lines are guides to the eye.}
		\label{fig:fig3}
\end{figure}

In order to summarize the effect of the surrounding dielectric environment on the relative energy of the neutral and charged dark excitons in reference to the bright one, we present the relative energy as a function of the average dielectric constant of the surrounding media, $\bar{\varepsilon}=(\varepsilon_t+\varepsilon_b)/2$. 
We define $\varepsilon_t$ and $\varepsilon_b$ as the relative dielectric constants of the top and bottom layers embedding the ML, respectively (in units of the vacuum permittivity, $\varepsilon_0$). 
We used the high-frequency values for the various dielectric constants due to the energy scale of the exciton binding energy~\cite{Stier2016}. 
The following dielectric constants for different media were applied: hBN flakes - $\varepsilon_{hBN}$=4.5, helium exchange gas - $\varepsilon_{He}$=1, and SiO$_2$ layer - $\varepsilon_{SiO_2}$=2.1. 
Fig.~\ref{fig:fig3} presents the comparison of our results devoted to the relative energies of the neutral and charged dark excitons in reference to the bright ones with those reported in the literature~\cite{Molas2017, Zhang2017, Robert2017, Wang2017, Barbone2018, Chen2018, Li2018, Paur2019, Molas2019, LiuGate, Li2019, arora2019dark, LiuValley, Liu2020, He2020, Sven2020, jadczak2021, yang2022}. 
It is seen that the reported relative energies of the X$^\textrm{D}$ and T$^\textrm{D}$ lines for WSe$_2$ ML encapsulated in hBN ($\kappa$=4.5) are distributed over a broad energy range of about 6-7~meV with the mean values centered at 42~meV and 58~meV, respectively. 
This suggests the presence of extra perturbations, $e.g.$ strain, which analysis stays beyond the scope of this work. 
The reduction of $\bar{\varepsilon}$ to the value of 1.6 results in a small increase of the corresponding relative energies to about 47 meV for the X$^\textrm{D}$ and 64 meV for the T$^\textrm{D}$. 
It has been recently demonstrated in Ref.~\citenum{Kapuscinski2021} that the magnitude of the 2D screening length and the surrounding dielectric screening is the same for the binding energies of both the neutral bright and dark excitons in the WSe$_2$ ML encapsulated in hBN. 
As a result, the variation of the average dielectric constant influences similarly the energies of both the X$^\textrm{D}$ and X$^\textrm{B}$ lines. 

In conclusion, the effect of the surrounding dielectric environment on the emission of the neutral and charged dark excitons has been investigated. 
The brightening coefficients of the dark excitons can be affected by the level of carrier concentration in the ML. 
Moreover, the surrounding media, characterized by different dielectric constants, influence weakly the relative energies of the neutral and charged dark excitons in reference to the bright ones.

\section*{Supplementary Material}

See the Supplementary Material for low-temperature PL spectra of the WSe$_2$ monolayer, PL spectra of the WSe$_2$ ML as a function of in-plane magnetic field, and estimation of carrier concentration in the WSe$_2$ ML.

\begin{acknowledgments}
The work has been supported by the National Science Centre, Poland (grant no. 2017/27/B/ST3/00205 and 2018/31/B/ST3/02111) and the CNRS via IRP "2DM" project. K. W. and T. T. acknowledge support from the Elemental Strategy Initiative conducted by the MEXT, Japan (grant no. JPMXP0112101001), JSPS KAKENHI (grant no. JP20H00354), and the CREST (JPMJCR15F3), JST. P. K. acknowledges support from the ATOMOPTO Project (TEAM programme of the Foundation for Polish Science, co-financed by the EU within the ERD-Fund).
\end{acknowledgments}

\section*{Data Availability Statement}
The data that support the findings of this study are available from the corresponding author upon reasonable request. 

\subsection*{}
The following article has been accepted by Applied Physics Letters. After it is published, it will be found at \href{https://dx.doi.org/10.1063/5.0085950}{https://dx.doi.org/10.1063/5.0085950}.

\bibliographystyle{apsrev4-1}
\bibliography{biblio}

\onecolumngrid
\newpage
\setcounter{figure}{0}
\setcounter{section}{0}
\renewcommand{\thefigure}{S\arabic{figure}}
\renewcommand{\thesection}{S\arabic{section}}

\begin{center}
	%%%%%%%%% ABSTRACT TITLE
	{\large{ {\bf Supplementary Material\\The effect of dielectric environment on the brightening of neutral and charged dark excitons in WSe$_2$ monolayer}}}
	%%%%%%%%% ABSTRACT AUTHORS
	\vskip0.5\baselineskip{Ma\l{}gorzata Zinkiewicz,{$^{1}$} Magdalena Grzeszczyk,{$^{1}$} \L{}ucja Kipczak,{$^{1}$} Tomasz Kazimierczuk,{$^{1}$} Kenji Watanabe,{$^{2}$} Takashi Taniguchi,{$^{3}$} Piotr Kossacki,{$^{1}$} Adam Babi\'nski,{$^{1}$} and Maciej R. Molas {$^{1}$}}
	
	%%%%%%%%% AFFILIATION
	\vskip0.5\baselineskip{\em$^{1}$ Institute of Experimental Physics, Faculty of Physics, University of Warsaw, ul. Pasteura 5, 02-093 Warsaw, Poland \\$^{2}$ Research Center for Functional Materials, National Institute for Materials Science, 1-1 Namiki, Tsukuba 305-0044, Japan \\$^{3}$ International Center for Materials Nanoarchitectonics, National Institute for Materials Science, 1-1 Namiki, Tsukuba 305-0044, Japan}
\end{center}

\section{Low-temperature PL spectra of the WSe$_2$ monolayer \label{sec:PL}}

\begin{figure*}[h!b]
	%	\subfloat{}%
	\centering
	\includegraphics[width=1\linewidth]{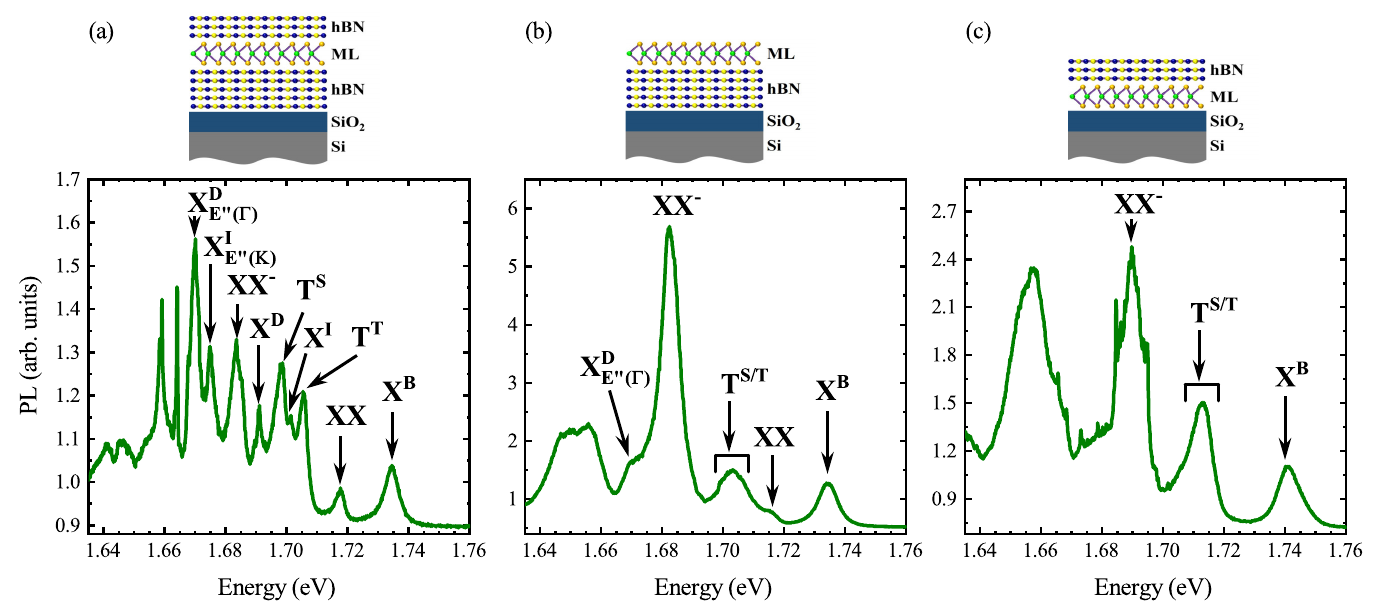}%
	\caption{PL spectra of WSe$_2$ monolayer: (a) encapsulated in hBN flakes, (b) deposited on hBN flake, and (c) deposited on SiO$_2$/Si substrate and covered with hBN flake at $T$=10~K measured at zero field. The PL spectra were normalized to the intensity of the bright X$^\textrm{B}$ line. }
	\label{fig:figS1}
\end{figure*}

Fig.~\ref{fig:figS1} demonstrates the photoluminescence (PL) spectra measured on a WSe$_2$ monolayer (ML) encapsulated in hBN flakes, deposited on hBN flake, and deposited on SiO$_2$/Si substrate and covered with hBN flake. The PL spectrum of the WSe$_2$ ML embedded in between hBN flakes displays several emission lines with a similar characteristic pattern already reported in a number of previous works on WSe$_2$ monolayers~\cite{Arora2015, Smolenski2016, Molas2017, Zhang2017, Robert2017, Wang2017, Barbone2018, Chen2018, Li2018, Koperski2019, Paur2019, Molas2019, LiuGate, Li2019, arora2019dark, LiuValley, Liu2020, He2020, Sven2020,jadczak2021, yang2022}. In accordance with these reports, the assignment of the observed emission lines is as follows: X$^\textrm{B}$ — a bright neutral exciton formed in the vicinity of the A exciton; XX - a neutral biexciton; T$^\textrm{S}$ and T$^\textrm{T}$ - singlet (intravalley) and triplet (intervalley) negatively charged excitons, respectively; X$^\textrm{I}$ and X$^\textrm{D}$ - momentum- and spin-forbidden dark excitons, respectively; XX$^-$ - a negatively charged biexciton; X$^\textrm{D}_\textrm{E"(K)}$ and X$^\textrm{D}_{\textrm{E"}(\Gamma)}$ - phonon replicas of the dark excitons associated with emissions of E" phonons from K and $\Gamma$ points of the Brillouin zone, respectively. For other MLs, $i.e.$ deposited on hBN flake, deposited on SiO$_2$/Si substrate and covered with hBN flake, the knowledge of the assignment of emission lines is limited and we were able to ascribe them to a few complexes. 

\section{PL spectra of the WSe$_2$ ML as a function of in-plane magnetic field \label{sec:field}}

\begin{figure*}[h!t]
	%	\subfloat{}%
	\centering
	\includegraphics[width=1\linewidth]{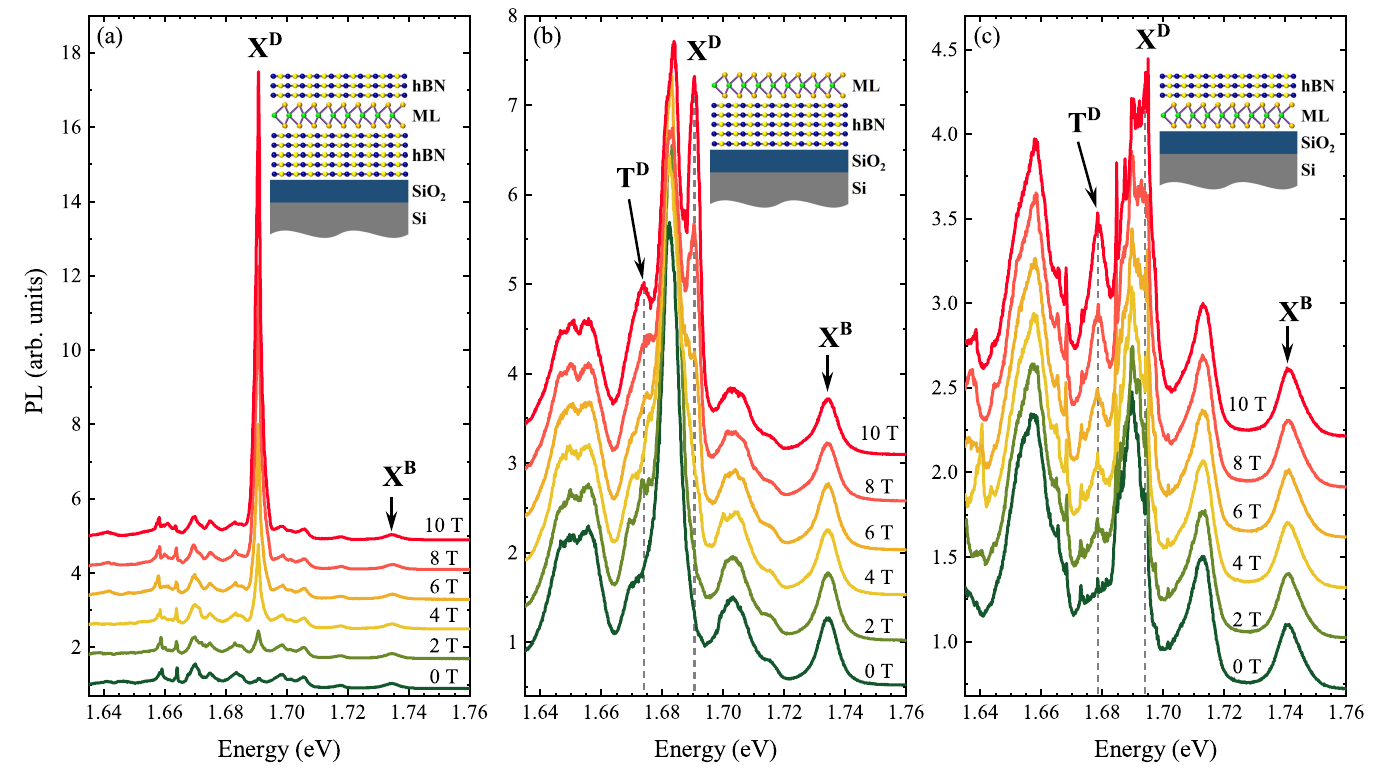}%
	\caption{Low-temperature ($T$=10~K) PL spectra of WSe$_2$ monolayer: (a) encapsulated in hBN flakes, (b) deposited on hBN flake, and (c) deposited on SiO$_2$/Si substrate and covered with hBN flake measured at the selected values of the in-plane magnetic fields. The PL spectra were normalized to the intensity of the bright X$^\textrm{B}$ line and shifted vertically for clarity. }
	\label{fig:figS2}
\end{figure*}

Fig.~\ref{fig:figS2} presents the PL spectra measured on a WSe$_2$ monolayer (ML) encapsulated in hBN flakes, deposited on hBN flake, and deposited on SiO$_2$/Si substrate and covered with hBN flake measured at selected values of the in-plane magnetic field. As can be appreciated in the figure, the brightening effect of the in-plane magnetic field is the largest for the ML encapsulated in hBN flakes. The X$^\textrm{D}$ intensity is several times bigger than the X$^\textrm{B}$ one. For other MLs, $i.e.$ deposited on on hBN flake, deposited on SiO$_2$/Si substrate and covered with hBN flake, the analogous effect is less significant. Nevertheless, the X$^\textrm{D}$ and T$^\textrm{D}$ can be well resolved at the highest values of the applied in-plane magnetic field.

\section{Estimation of carrier concentration in the WSe$_2$ ML \label{sec:comparison}}

\begin{figure*}[h!t]
	%	\subfloat{}%
	\centering
	\includegraphics[width=1\linewidth]{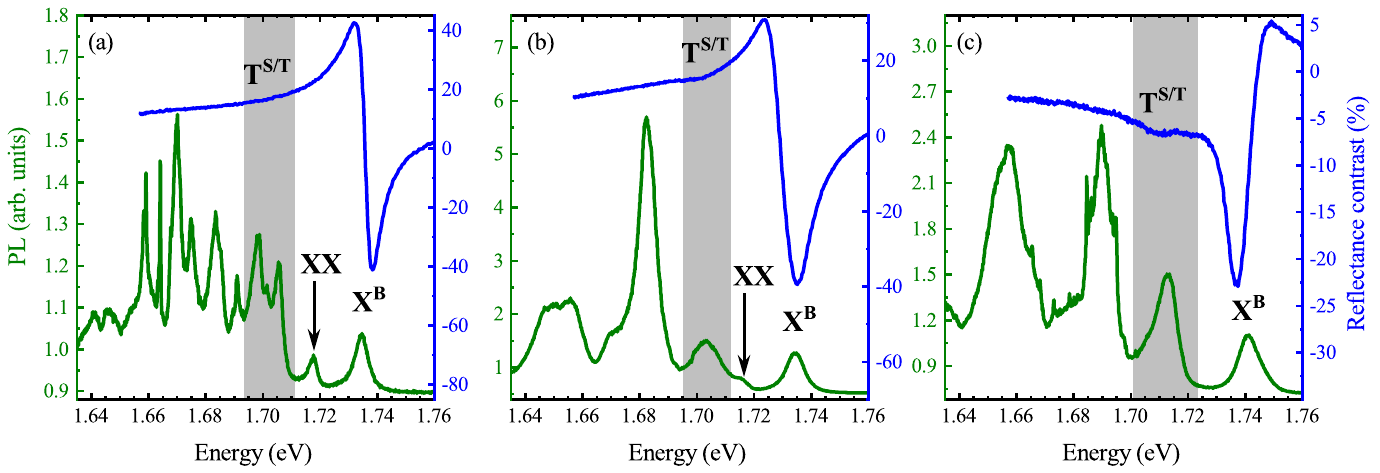}%
	\caption{Low-temperature ($T$=10~K) PL and RC spectra of WSe$_2$ monolayer: (a) encapsulated in hBN flakes, (b) deposited on hBN flake, and (c) deposited on SiO$_2$/Si substrate and covered with hbN flake measured at the selected values of the in-plane magnetic fields. The PL spectra were normalized to the intensity of the bright X$^\textrm{B}$ line. The gray shadowed regions denotes the energy ranges of the emission due to the negatively charged excitons.}
	\label{fig:figS3}
\end{figure*}

It is proposed in the manuscript that the observed significant change of the brightening effect of the X$^\textrm{D}$ and T$^\textrm{D}$ line measured on the ML in different dielectric environments is due to the variation of the electron concentration. 
The Fermi energy ($E_F$) can be evaluated using the formula $\Delta E$=$E_{\textrm{X}^\textrm{B}}-E_{\textrm{T}^\textrm{S/T}}=E_b+E_F$, where $E_{\textrm{X}^\textrm{B}}$ and $E_{\textrm{T}^\textrm{S/T}}$ are energies of the X$^\textrm{B}$ and T$^\textrm{S/T}$ lines, respectively, and $E_b$ corresponds to the binding energy of the negative trion. 
The binding energies of both the spin-singlet and spin-triplet trions can be obtained from the measurements of the gated WSe$_2$ monolayer. 
Based on the literature~\cite{Courtade2017,LiuGate}, we obtained that the binding energies for the T$^\textrm{T}$ and T$^\textrm{S}$ trions are of about 28.3 meV and 35.0 meV, respectively. 
The $\Delta E$ values were found to be on the order of 29.1 meV ($E_{\textrm{X}^\textrm{B}}-E_{\textrm{T}^\textrm{T}}$) and 36.3 meV ($E_{\textrm{X}^\textrm{B}}-E_{\textrm{T}^\textrm{S}}$) for the results presented in Fig.~\ref{fig:figS3}(a), which imply the corresponding Fermi energies equals to about 0.8 meV and 1.3 meV. Using the relation $n=m_e E_F/\pi \hbar^2$ and assuming the effective mass of electrons $m_e$=0.4\cite{Kormanyos2015}, we can estimate the free electron concentration to be of the order of $n\sim 1.3-2.2 \times 10^11$ cm$^-1$. At the same time, the corresponding RC spectrum, defined as $RC(E)=(R(E)-R_0(E))/(R(E)+R_0(E))\times100\%$ ($R(E)$ and $R_0(E)$ are the reflectance of the investigated ML and of the same structure without the WSe$_2$ monolayer, respectively), does not comprise signal in the energy range of charged excitons. It allows us to conclude that the ML encapsulated in hBN flakes is close to the neutrality point. Unfortunately, the similar analysis of the Fermi energy can not be performed on two other regions of the WSe$_2$ ML due to the lack of the PL measurements as a function of gate voltage for MLs in the particular dielectric environment. However, the doping level can be evaluated qualitatively for these two regions. As there are small resonances observed in the RC spectra (see Fig.~\ref{fig:figS3}(b) and (c)), which energies coincide with the T$^\textrm{S/T}$ line, we can conclude that the free electron concentration is non-negligible in these two cases. Moreover, the emission line attributed to the neutral biexciton (XX) is apparent in the PL spectrum of the WSe$_2$ ML deposited on hBN flakes, while is absent for the WSe$_2$ ML deposited on SiO$_2$/Si substrate and covered with hBN flake. It suggests that the free electron concentration is the highest in the ML part deposited on SiO$_2$/Si substrate and covered with hBN flake. Consequently, we can conclude that the doping level of the ML encapsulated in hBN is the smallest, it increases with the ML on hBN flake, leaving the ML deposited on SiO$_2$/Si and covered with hBN flake with the highest carrier concentration. 

\bibliographystyle{apsrev4-1}
\bibliography{biblio}

\end{document}